\documentclass{moriond}

\usepackage{epsfig,graphics}

\newcommand\apj{ApJ}
\newcommand\apjs{ApJS}
\newcommand\mnras{MNRAS}
\newcommand\prd{Phys.~Rev.~D}

\def\be{\begin{equation}}
\def\ee{\end{equation}}
\def\bea{\begin{eqnarray}}
\def\eea{\end{eqnarray}}

\newcommand{\Photo}{}  

\begin{document}
\vspace*{4cm}
\title{Massive Neutrinos, Dark Sector, and Hydrodynamics: The \textit{Sejong} Suite}
\author{GRAZIANO ROSSI}
\address{Department of Physics and Astronomy, Sejong University, Seoul, 143-747, Korea}



\maketitle\abstracts{
Improving  the understanding of the processes governing small-scale nonlinear clustering
is a necessary task for interpreting upcoming high-quality cosmological data, as it will allow one to 
break degeneracies and obtain tight neutrino mass and warm dark matter constraints from
large-scale structure probes. The \textit{Sejong Suite}, 
an extensive collection of 
state-of-the-art high-resolution cosmological hydrodynamical simulations,
has been intended with this primary goal in mind.
Spanning a large number of cosmological and astrophysical parameters
(especially suitable for the \textit{dark sector}), 
and organized into three
main categories (\textit{Grid Suite}, \textit{Supporting Suite}, and \textit{Systematics Suite}), the 
release  may be useful for a broader variety of cosmological and astrophysical purposes -- while 
primarily developed for Lyman-$\alpha$ (Ly$\alpha$) forest studies. 
In particular, the overall architecture
of the \textit{Grid Suite} has been designed to achieve an equivalent resolution 
up to $3 \times 3328^3 = 110$ billion particles in a ($100h^{-1}{\rm Mpc}$) box, corresponding to a $30h^{-1}{\rm kpc}$ 
mean grid resolution, which ensures convergence on Ly$\alpha$ flux statistics closer to the desired $\sim 1.0\%$ level that 
data from surveys such as the  Dark Energy Spectroscopic Instrument (DESI) will provide. 
Here, we briefly highlight the main 
characteristics, improvements, and novelties of the  \textit{Sejong Suite}, as well as ongoing and future applications.} 



\section{Scientific Rationale: Dark Sector Cosmology at Small Scales}


The ability to reach small nonlinear scales  will be crucial in the next few years, as
it will allow one to break degeneracies and contribute to tightening neutrino mass,
dark radiation, and warm dark matter (WDM) constraints derived from large-scale structure (LSS) tracers.
Small nonlinear scales are in fact the key for competitive massive neutrinos and \textit{dark sector} (i.e., dark radiation, WDM) bounds, and in
this regards the 
remarkable potential of the Lyman-$\alpha$ (Ly$\alpha$) forest resides in 
the ability to reach such regime, still inaccessible to other probes.
The Ly$\alpha$ forest is highly sensitive to the summed neutrino mass ($M_{\nu}$) and additional dark radiation components such as sterile neutrinos 
(i.e., when the effective number of neutrino species $N_{\rm eff}$ departs from its canonical value), via significant attenuation effects on the matter and flux power spectra at small scales. 
\cite{Sel2005,Rossi2015a,Rossi2017} 
Currently, neutrino mass upper bounds from cosmology 
 are closer to the minimum value allowed by the inverted hierarchy (IH), and   the  Ly$\alpha$  forest in synergy 
 with cosmic microwave background (CMB) data provides among the strongest 
reported constraints on $M_{\nu}$ and $N_{\rm eff}$.\cite{Sel2005,Rossi2015b,Planck2020,eBOSS2021} 
State-of-the-art surveys such as the Extended
Baryon Oscillation Spectroscopic Survey (eBOSS)
are already highly competitive in this regards \cite{eBOSS2021}, but 
ongoing experiments like the  Dark Energy Spectroscopic Instrument (DESI) \cite{DESI2016}
are expected to greatly improve on current \textit{dark sector} bounds 
in terms of statistical precision and accuracy.
Note also that 
conclusions regarding the minimal six-parameter $\Lambda$CDM concordance 
cosmological scenario dominated by cold dark matter (CDM) and a dark energy (DE) component in the form of a cosmological constant $\Lambda$
are mainly inferred from LSS
observations -- while small scales still remain poorly investigated.
In addition, while the role of the Ly$\alpha$ forest is critical in sharpening 
\textit{dark sector} constraints, current results are heavily based on numerics and on details related to statistical analysis methodologies. 
Advancement in the modeling and a deeper understanding of small neutrino mass effects on key Ly$\alpha$
observables (particularly on the transmitted flux power spectrum) and of possible systematics (i.e., impact of complex 
small-scale baryonic physics) are needed to enhance the robustness of all  Ly$\alpha$-based studies. 
Moreover, progress in the modeling of systematics, ameliorations in the small-scale modeling, and a better theoretical understanding of neutrino 
mass and dark radiation effects on cosmological observables are necessary for a reliable use of 
LSS data to robustly constrain $M_{\nu}$ and $N_{\rm eff}$. 
All these considerations represent the primary motivations and  scientific rationale of our work. 



\section{The \textit{Sejong} Hydrodynamical Simulation Suite: Highlights}

 
\begin{figure}[tb]
\centering
\includegraphics[width=140mm]{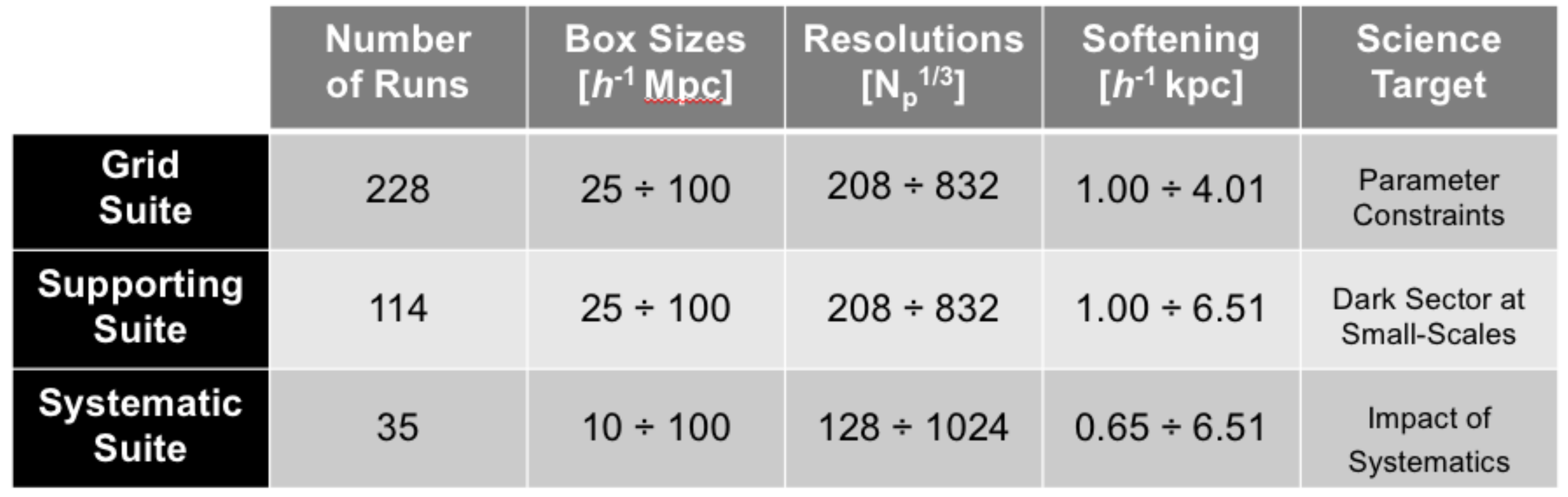}
\caption{\textit{Sejong Suite}: selected technical details.}
\label{tab1}
\end{figure}

\begin{figure}[tb]
\centering
\includegraphics[width=150mm]{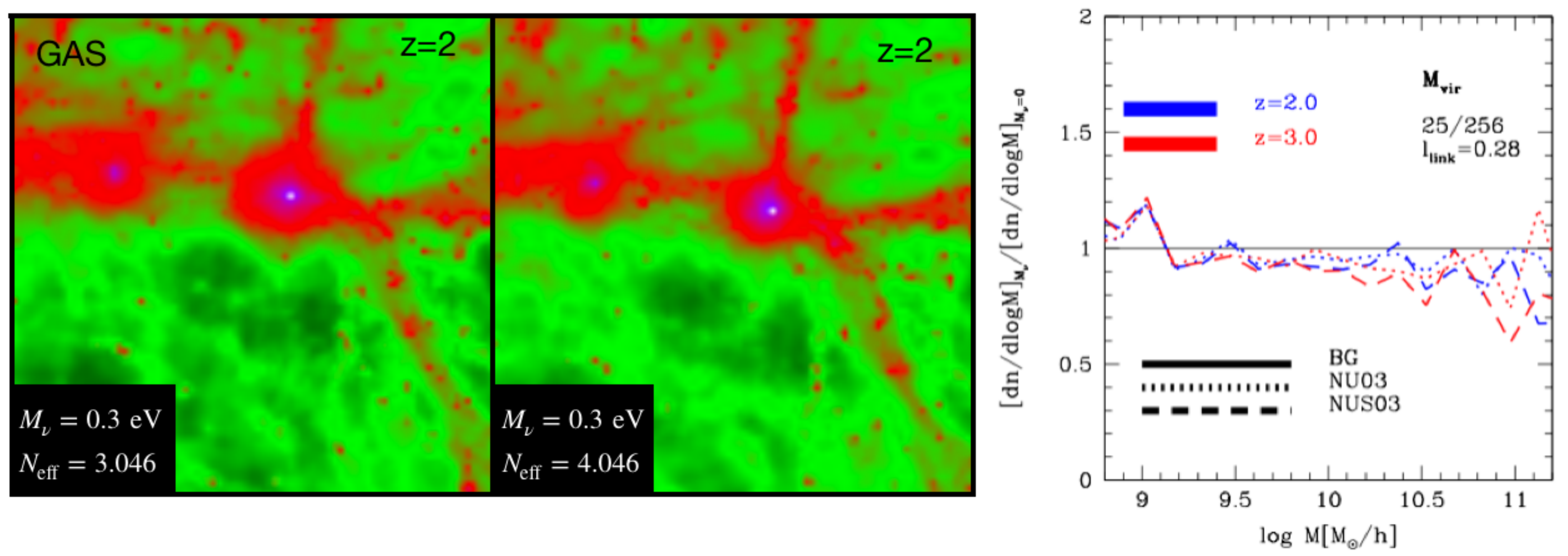}
\caption{[Left] $4\times4 h^{-1}{\rm Mpc}$ patch at $z=2$ in a massive neutrino cosmology with
$M_{\nu} =0.3{\rm eV}$, and in a mixed scenario where also a massless sterile neutrino 
is added (i.e., $N_{\rm eff} = 4.046$). [Right]  Relative halo abundances  
at $z=2$ and $z=3$ in the corresponding models, normalized by the baseline massless neutrino cosmology.}
\label{fig2}
\end{figure}
\begin{figure}[tb]
\centering
\includegraphics[width=150mm]{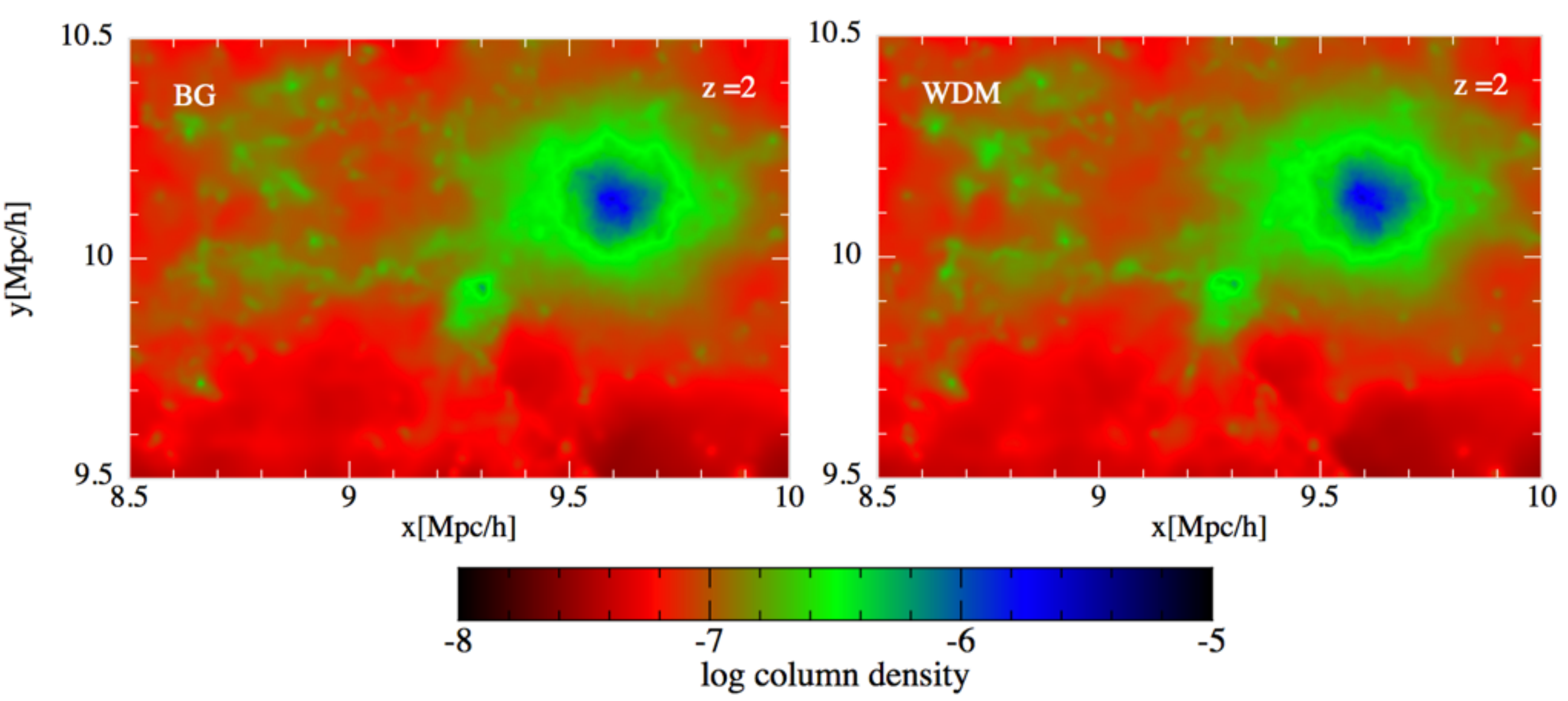}
\caption{Small projected patch at $z=2$ from $25h^{-1}{\rm Mpc}$ simulations with $256^3$ particles/type. 
A complex structure is displayed, as seen in the gas column density of the reference cosmology (left panel),
and in a WDM scenario when $m_{\rm WDM} = 2.00$ keV (right panel).
Although differences are hardly perceptible, the impact of a $m_{\rm WDM} = 2.00$ keV relic on the high-$z$ cosmic web is significant.}
\label{fig3}
\end{figure}

The  \textit{Sejong Suite} \cite{Rossi2020}  is organized into three main categories, 
addressing different scientific and technical aspects.
The \textit{Grid Suite} ($76 \times 3 = 228$ simulations)
is targeted primarily for cosmological parameter constraints, with a primary focus on   
massive and sterile neutrinos and the \textit{dark sector}, exploiting the small-scale 
flux power spectrum; it represents our leading effort and major deliverable.
The \textit{Supporting Suite} (114 simulations) is aimed at studying the detailed physical 
effects of exotic particles and dark radiation models, as well as their impact on the high-$z$ cosmic web.
The \textit{Systematics Suite} (35 realizations) is useful for 
quantifying several systematic effects, ranging from numerical challenges until parameter degeneracies. 
Table \ref{tab1} summarizes a number of key characteristics of the three groups.  
The simulations have been produced at the Korea Institute of Science and Technology Information (KISTI) supercomputing infrastructure,
and are based on a modified version of \texttt{Gadget-3} \cite{Springel2005} used
for evolving Euler hydrodynamical equations and primordial chemistry. 
The pipeline is interfaced with \texttt{CAMB} \cite{Camb2000} and a modified version of second-order Lagrangian perturbation theory (2LPT). 
Primarily, the suite is targeted to explore the high-$z$ 
cosmic web as seen in the Ly$\alpha$ forest ($5.0 \le z \le 2.00$), although all of the realizations may have a much broader usage. 
The release contains a number of improvements, in relation to  
efficiency of the pipeline, resolution,
grid and stepsize accuracy, cosmology, and reionization history. Full details are provided in {\color{blue} Rossi (2020)}. 
The most interesting novelty is the presence of  mixed models, including the
combined effects of WDM, neutrinos, and dark radiation. 
An example is provided in Figure \ref{fig2}. The left panel show
a small $4\times4 h^{-1}{\rm Mpc}$ patch at $z=2$ in a massive neutrino cosmology with
$M_{\nu} =0.3{\rm eV}$, as well as in a mixed scenario where also a massless sterile neutrino 
is added (i.e., $N_{\rm eff} = 4.046$). 
Differences in the cosmic web morphology, 
albeit small, are clearly visible. The right panel shows  
the relative halo abundances  
at $z=2$ and $z=3$, respectively, in the corresponding models, normalized by the baseline massless neutrino cosmology.
Halos are extracted with the \texttt{Rockstar} \cite{Behroozi2013} algorithm, assuming a linking length $l_{\rm link}=0.28$. 
In essence, the presence of massive neutrinos and dark radiation delays  structure formation:
this  primarily affects the higher end of the mass function, which is modified  depending on redshift and neutrino 
mass and/or $N_{\rm eff}$ -- because of free-streaming effects at small scales. 
As in our previous releases, we adopt a particle-based implementation of massive neutrinos.  
Regarding dark radiation,  we consider models with 4 neutrinos, 
where 3 are massive and active while the fourth one is massless, sterile, and thermalized with the active ones -- so that $N_{\rm eff} = 4.046$.
As far as WDM, we focus on two implementations, following different methodologies. 
In both cases, we only consider early decoupled thermal relics, and assume that all DM is warm when massive neutrinos are not present (i.e., pure WDM models);
suitable candidates are keV right-handed neutrinos or sterile neutrinos. 
As an example, Figure \ref{fig3} shows a small projected patch from simulations with a box size of $25h^{-1}{\rm Mpc}$ and $256^3$ particles/type, describing the gas
column density at $z = 2.0$. The left panel displays a complex structure as seen in the reference model, while the right panel highlights the same structure as seen in a WDM 
cosmology when $m_{\rm WDM} = 2.00$ keV. Once again, differences are tiny and therefore hardly detectable, although the impact of a $m_{\rm WDM} = 2.00$ keV relic on the high-$z$ LSS is not negligible.


\section{Contributions, Applications, Outlook}


Reaching a very high sensitivity on small scales and resolving baryonic physics are essential aspects for improving $M_{\nu}$, $N_{\rm eff}$, 
and \textit{dark sector} bounds, as well as for breaking degeneracies -- as demanded by upcoming high-quality data.
The addition of accurate small-scale observations, soon available, will in fact allow one to break degeneracies, and hence contribute to tighten
neutrino mass and \textit{dark sector} constraints from LSS probes. Motivated by these goals, we 
have carried out an extensive set of 
high-resolution cosmological hydrodynamical simulations (over 300 runs) termed the \textit{Sejong Suite}, 
primarily developed for modeling the Ly$\alpha$ forest in the redshift interval $5.0 \le z \le 2.0$  -- although the simulations may have a much broader usage. 
In particular, the overall design has been targeted to meet the demanding resolution of the expected DESI Ly$\alpha$ forest data (i.e., $30h^{-1}{\rm kpc}$ 
mean grid resolution). The \textit{Sejong Suite} features a number of improvements and novelties at all levels 
with respect to our previous releases, in relation to technical, modeling, and innovative aspects.
Noticeably, we have expanded the parameter space for the \textit{Grid Suite} and tighten their variation range. 
We have also addressed a series of nontrivial systematics,
and  produced more than 288 million Ly$\alpha$ forest skewers mapping an extended parameter space.
On the innovative side, the most significant novelty is the inclusion, for the first time, of extended mixed scenarios describing the combined effects of WDM, neutrinos, and dark radiation. 
These non-canonical models are quite interesting, particularly for constraining $N_{\rm eff}$ and WDM relic masses directly from Ly$\alpha$ forest observations.
Our work is thus useful for interpreting upcoming high-quality data from eBOSS and DESI (with ongoing data applications), and synergetic to particle physics experiments.
In future releases of the \textit{Sejong Suite}, we plan to expand around this baseline framework and provide more refined realizations.



\section*{Acknowledgments}
G.R. is supported by the National Research Foundation of Korea (NRF) through Grant No. 2020R1A2C1005655 funded by the Korean Ministry of Education, 
Science and Technology (MoEST). The numerical simulations presented in this work were performed using the Korea Institute of Science and Technology Information (KISTI) 
supercomputing infrastructure (Tachyon 2) under allocations KSC-2017-G2-0008 and KSC-2018-G3-0008, 
and post-processed with the KISTI KAT System (KISTI/TESLA `Skylake' and `Bigmem' architectures) under allocations KSC-2018-T1-0017, KSC-2018-T1-0033, and KSC-2018-T1-0061. 
We also acknowledge extensive usage of our new computing resources (Xeon Silver 4114 master node and Xeon Gold 6126 computing node architecture)  at Sejong University.



\section*{Selected References}


\end{document}